\documentclass[aps,amssymb,prl,twocolumn,showpacs,amsmath]{revtex4}
\usepackage{epsfig}
\usepackage{bm}
\usepackage{ulem}

\begin{document}

\title{Dark soliton dynamics in Bose-Einstein condensates at finite 
temperature}
\author{B.~Jackson}
\author{N.~P.~Proukakis}
\author{C.~F.~Barenghi}
\affiliation{School of Mathematics and Statistics, University of 
 Newcastle upon Tyne, NE1 7RU, United Kingdom}
          
\begin{abstract}
 The dynamics of a dark soliton in an elongated Bose-Einstein condensate 
 at finite temperatures is studied using numerical simulations. We find
 that in the presence of harmonic confinement the soliton may oscillate
 even at finite temperatures, but with an amplitude that increases with
 time, indicating the decay of the soliton.
 The timescale of this decay decreases both with  
 increasing temperature and with increasing initial soliton velocity. 
 Simulations performed for the experiment of S.~Burger 
 {\it et al.}, Phys.~Rev.~Lett.~{\bf 83}, 5198 (1999), reveal excellent
 agreement with the observed soliton decay, confirming the crucial role
 of the thermal cloud in soliton dynamics.
\end{abstract}
\date{\today}


\maketitle

Bose-Einstein condensates (BECs) in ultracold gases provide an ideal 
test ground for nonlinear physics. At temperatures close to absolute zero, the 
condensate can be accurately described by the Gross-Pitaevskii (GP) 
equation \cite{pitaevskii}, 
which has the form of a nonlinear Schr\"{o}dinger equation where the nonlinear
term arises from interactions between atoms. A similar equation
appears in nonlinear optics, where it is known to support soliton solutions
\cite{kivshar98}. It is natural, therefore, to expect solitons in BECs.

Both bright and dark solitons have indeed been observed in experiments. Bright 
solitons can be formed when the atomic interactions are
attractive \cite{strecker02,khaykovich02,cornish06}, whereas ``gap'' bright 
solitons have been created experimentally in repulsive condensates 
in optical lattices \cite{eiermann03}.
Repulsive condensates in harmonic traps, however, can only support dark 
solitons, which correspond to propagating one-dimensional localized minima in 
the density. 
Dark solitons have been generated experimentally
\cite{burger99,denschlag01,dutton01,anderson01}, and their zero 
temperature properties have been studied theoretically in both 
three-dimensional geometries, 
where they exhibit dynamical instabilities 
\cite{anderson01,feder00,muryshev99}, 
as well as in one dimension \cite{busch00,frantzeskakis02,parker03,konotop04}.

Although modelling the condensates at zero temperature provides important 
insight into dark solitons, in practice 
experiments are always performed at non-zero temperatures, where the presence 
of a cloud of non-condensed particles provides a mechanism for the decay of 
the soliton. Despite the experimental evidence for
such dissipative effects, the finite 
temperature properties of solitons have received 
relatively little attention. Refs.\ \cite{fedichev99,muryshev02}
predicted dissipation at finite temperatures by 
considering the reflection of excitations from a soliton in a uniform 
condensate. The role of 
quantum fluctuations has also been studied \cite{dziarmaga03}.

\begin{figure}[h]
\centering \scalebox{0.42}
 {\includegraphics{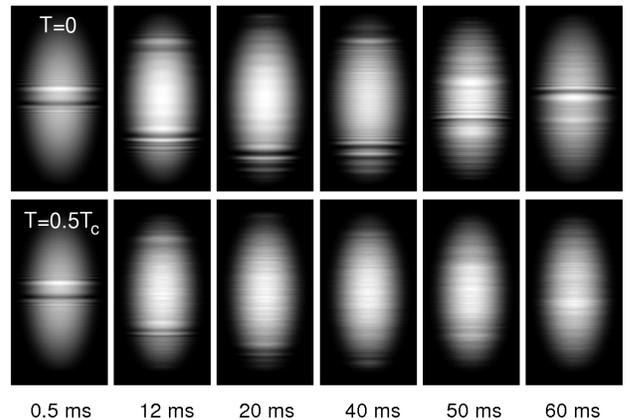}}
 \caption{Evolution of a dark soliton at 
 $T=0$ (top) and $T=0.5\, T_c$ (bottom), 
 corresponding to the experimental parameters of Ref.\ \cite{burger99}. In 
 each simulation we initially imprint a phase on the condensate, then allow
 it to evolve for the time shown before releasing it from the trap. Plotted 
 are condensate column densities after a subsequent expansion of 
 $t=4\, {\rm ms}$.} 
\label{fig:expcol}
\end{figure}

This Letter uses numerical simulations to perform a detailed quantitative 
study of the dissipative dynamics of dark 
solitons in elongated harmonic traps at finite temperatures. At zero 
temperature, a dark soliton is predicted to oscillate in the 
axial direction \cite{muryshev99,busch00,frantzeskakis02,parker03,konotop04}.
However, the presence of a thermal cloud leads to damping.
As a result, the depth of the propagating soliton decreases in time, 
leading to an increase in the amplitude of the oscillations, which eventually 
approaches the half-length of the condensate. 
This is consistent with a decrease in the soliton energy as a function
of time. We perform detailed investigations of this decay for 
different temperatures and initial 
soliton depths, and separately assess the role of mean field coupling and
binary collisions between the atoms. In particular, we directly simulate the 
experiment of Burger {\it et al.\ }\cite{burger99}, where the evolution of
dark solitons, created by phase imprinting, was observed using 
time-of-flight absorption imaging. Our simulations show that the soliton
effectively disappears after only half an oscillation (Fig.\ \ref{fig:expcol}, 
bottom images), in good agreement with the 
experimental findings, thus proving the crucial role of the thermal cloud. 
This should be contrasted with the undamped dynamics predicted by the GP 
equation (top images).
 
Our simulations are based on the formalism of Zaremba, Nikuni, and Griffin 
(ZNG) \cite{zaremba99}, where the dynamics of the condensate and the thermal 
cloud are described by the coupled equations:
\begin{equation}
 i\hbar \frac{\partial \Psi}{\partial t} = \left( -\frac{\hbar^2 \nabla^2}
 {2m} + V +gn_c + 2g\tilde{n} - iR \right) \Psi, 
\label{eq:GP-gen}
\end{equation}
\begin{equation}
 \frac{\partial f}{\partial t} + \frac{\bm p}{m} \cdot {\bm \nabla} f 
 -{\bm \nabla} U \cdot {\bm \nabla_p} f = C_{12} + C_{22}.  
\label{eq:Boltz}
\end{equation}
Eq.~(\ref{eq:GP-gen}) is a generalized GP equation for the 
condensate wavefunction $\Psi({\bm r}, t)$, while Eq.~(\ref{eq:Boltz}) is a 
Boltzmann equation for
the thermal cloud phase space density $f({\bm p},{\bm r},t)$ 
(where ${\bm p}$ is the momentum). The 
condensate and thermal cloud densities are defined as $n_c = |\Psi|^2$ and 
$\tilde{n}=\int f \,d^3p/h^3$ respectively, while
$g=4\pi \hbar^2 a/m$ parameterizes the mean field interactions between atoms of 
mass $m$ and scattering length $a$. The effective potential is 
$U=V+2g(n_c+\tilde{n})$, with $V=m(\omega_{\perp}^2 r^2+\omega_z^2 z^2)/2$ 
representing the harmonic trap. Apart from the 
mean field coupling between the condensate and non-condensate, Eq.\
(\ref{eq:GP-gen}) contains a term 
$R = (\hbar /2n_c) \int C_{12} \,d^3p/h^3$ arising from collisional processes,
$C_{12}$, that add or remove a single particle from
the condensate.  The $C_{22}$ term in Eq.\ (\ref{eq:Boltz}) denotes binary 
collisions involving only thermal atoms.

This theory has already been used to model the damping of collective modes, 
demonstrating good agreement with experiments (see e.g.\ Ref.\ 
\cite{jackson02b} and references therein), with the numerical methods used to 
solve these equations described in Ref.~\cite{jackson02a}, 
Here, we also model the dynamics of the thermal cloud 
using $N$ body simulations. Eq.~(\ref{eq:GP-gen}) is solved 
in cylindrical coordinates $(r,z)$ using a Crank-Nicholson scheme, with the
singularity in the Laplacian at $r=0$ avoided by offsetting the radial grid by 
half of a grid spacing.

For numerical convenience, our initial analysis is performed
for a relatively small sample of $N=2\times 10^4$ $^{87}{\rm Rb}$ atoms in a 
trap with 
frequencies $\omega_z = 2\pi\times 10\,{\rm Hz}$ and 
$\omega_{\perp}=2\pi\times 2500\,{\rm Hz}$. The cloud is thus highly 
elongated along the axial direction, with
the tight transverse confinement suppressing 
``snake-instabilities'' which lead to decay of the soliton into vortices
\cite{dutton01,anderson01,feder00,muryshev99}.

The initial condition for each simulation is established by first 
self-consistently
finding the equilibrium state of the condensate wavefunction and thermal 
cloud Bose distribution $f({\bm p},{\bm r})=\{ \exp[(p^2/2m+U-\mu)/k_B T]-1 
\}^{-1}$ (where $\mu$ is the condensate chemical potential) at a given
temperature.  A soliton is then generated by multiplying the ground state 
condensate wavefunction $\Psi({\bm r})$ with $\Psi_s (z) = \beta \tanh (\beta 
z/\xi) + i (v/c)$,
where $\beta=\sqrt{1-(v/c)^2}$, $\xi=\hbar/\sqrt{mgn_c}$ is the condensate 
healing length, and $c=\sqrt{gn_c/(2m)}$ is the speed of sound.
 
\begin{figure}[h]
\centering \scalebox{0.6}
 {\includegraphics{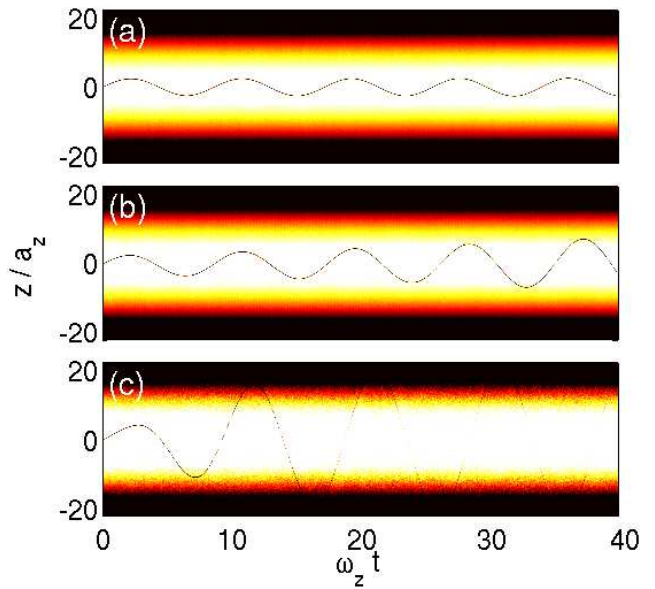}}
 \caption{(Color online) Temporal evolution of the condensate density as a 
 function of $z$ along a cross-section at $r=\Delta r/2$, where $\Delta r$ 
 is the grid spacing. Light colors represent high densities and 
 dark low densities, with the soliton appearing as a dark line.
 Both the position and time units are expressed in terms of the 
 axial trap frequency $\omega_z$, where $a_z = [\hbar/(m\omega_z)]^{1/2}$. The 
 three different plots are for (a) $T =0$, (b) $T\simeq 0.2\, T_c$, and
 (c) $T\simeq 0.4\,T_c$, where $T_c \simeq 500\, {\rm nK}$, and the initial 
 velocity is $v=0.1c$.}
\label{fig:soltog}
\end{figure}

Firstly, we consider only the effect of mean-field coupling ($n_c$,
$\tilde{n}$) between
the condensate and thermal cloud, setting $R=C_{12}=C_{22}=0$ in 
Eq.\ (\ref{eq:Boltz}). We perform simulations 
for a range of initial velocities $v$, where smaller $v$ is associated with 
deeper solitons (with a completely dark soliton being stationary). To study
the soliton decay process in detail, we first consider a rather
idealized example of a deep (slow) soliton with $v=0.1c$. The 
subsequent dynamics is illustrated in Fig.\ \ref{fig:soltog}, which shows 
the evolution of the longitudinal density profile for three different 
scaled temperatures, $T/T_c$, where $T_c$ is the critical temperature for 
BEC. The density inhomogeneity due to the harmonic trap is evident, with 
the soliton appearing as a dark line due to its low density.

Fig.\ \ref{fig:soltog}(a) illustrates the dynamics at $T=0$, where the soliton
oscillation has a constant amplitude and a frequency close to 
$\omega_z/\sqrt{2}$, in agreement with the predictions
of the GP equation \cite{muryshev99,busch00,frantzeskakis02}. In this case, 
the soliton maintains an almost 
constant depth, and the extrema of the oscillation occur when the density at 
the soliton minimum reaches zero. Fig.\ \ref{fig:soltog}(b) shows the
result at $T=100\, {\rm nK}$ ($0.2\,T_c$), where a small thermal 
cloud is also present. The steady increase in the oscillation 
amplitude accompanies a decrease in the soliton depth. At 
$T=200\, {\rm nK}$ ($0.4\,T_c$) [Fig.\ \ref{fig:soltog}(c)] the rate of 
increase in amplitude is larger, and the soliton quickly becomes very shallow,
making it difficult to visualize in the image.

This increase in oscillation amplitude is a consequence of damping of the 
soliton due to dissipation. This is most clearly seen by considering the 
energy of the soliton 
\begin{equation}
 E_s (r,t)= \int^{z_s+\delta}_{z_s-\delta} dz \, \left [ \frac{\hbar^2}{2m} 
 \left | \frac{d\Psi}{dz}\right |^2 + \frac{g}{2} (n_c-n_0)^2 \right ],  
\end{equation}
evaluated at $r=\Delta r/2$, which is the nearest grid 
point to the $z$-axis. The integral is taken over a small interval $2\delta$
centered on the soliton position, $z_s$, where we choose 
$\delta = 5\xi$. The background density for a condensate without a soliton is
represented by $n_0$. 

\begin{figure}[h]
\centering \scalebox{0.42}
 {\includegraphics{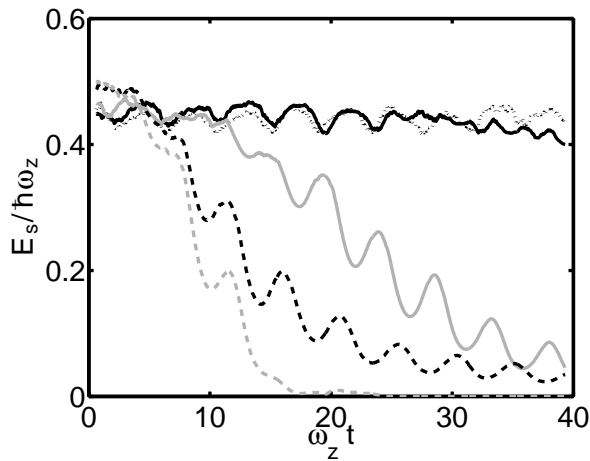}}
 \caption{Soliton energy, $E_s$, as a function of time for the parameters
 of Fig.\ \ref{fig:soltog}, and temperatures $T=0$ (dotted), 
 $T \simeq 0.2\,T_c$ (solid) and $T \simeq 0.4\,T_c$ (dashed). In the latter 
 two cases, black lines plot the collisionless simulations, while gray lines 
 include collisions.} 
\label{fig:sol_energy}
\end{figure}

In Fig.\ \ref{fig:sol_energy} we plot the soliton energy as a function of
time. We see that for $T=0$ (dotted black line, top) the energy is constant, 
apart from an oscillation arising from continuous soliton-sound interactions
\cite{parker03}: an accelerating soliton in a harmonic trap periodically emits
sound, leading to an instantaneous loss in energy. However, since the 
condensate is of finite size, the sound
cannot escape from the system and is therefore periodically re-absorbed by the
soliton, thus leading to no net decay in energy. In contrast, at
finite temperatures there is a gradual loss of energy. While 
this effect is only marginal at relatively low temperatures 
($T \simeq 0.2\,T_c$) represented by the black solid line, the dissipation is 
significantly enhanced with increasing temperature, as can be seen from 
the black dashed line ($T \simeq 0.4\,T_c$).
 
Although the collisionless regime discussed thus far captures some of the 
essential physics,
experiments with solitons are actually performed in the presence of 
collisions ($C_{12} \neq 0$, $C_{22} \neq 0$), to which we now turn our 
attention.
The energy decay when collisions are included is represented by the gray
lines in Fig.\ \ref{fig:sol_energy}. Comparing to the collisionless results,
one sees that the decay rate is greatly enhanced, and for $T \simeq 0.4\,T_c$
the energy falls to zero already at $\omega_z t \simeq 20$.

\begin{figure}[h]
\centering \scalebox{0.45}
 {\includegraphics{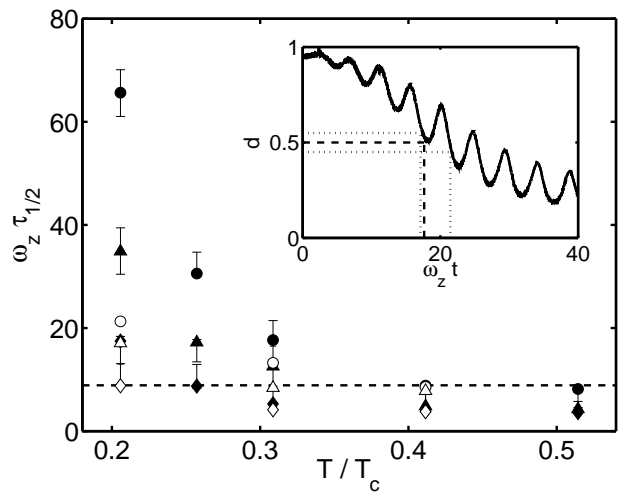}}
 \caption{Time, $\tau_{1/2}$, for the soliton depth to decay to half of its 
 initial value as a function of scaled temperature, $T/T_c$, for different
 initial velocities $v=0.1c$ (circles), $v=0.25c$ (triangles), and $v=0.5c$
 (diamonds). Solid (open) symbols correspond to collisionless
 (collisional) simulations. The dashed line marks the period of one soliton
 oscillation, $\omega_z t = 2^{3/2} \pi$. {\it Inset:} Time-dependence of the
 soliton depth, $d$, scaled to its initial
 value, for $T \simeq 0.3\, T_c$, $v=0.1c$, and $C_{12}=C_{22}=0$. This
 yields $\tau_{1/2}$ data points for the main graph (dashed lines)
 and the asymmetrical error bars obtained from $0.45<d<0.55$ (dotted lines).} 
\label{fig:halflife-ins}
\end{figure}

Simulations for the soliton dynamics have been performed for a range of 
temperatures, initial velocities, and with and without collisions.
In order to quantify the resulting soliton decay, we define a 
``half-life'', $\tau_{1/2}$, which corresponds to the time required for the 
soliton depth to reach half of its initial value. This quantity is relevant
since it is closely related to the contrast of the absorption images obtained 
experimentally after time-of-flight expansion.
The dependence of the half-life on the above parameters
is shown in Fig.\ \ref{fig:halflife-ins}. The procedure 
for obtaining $\tau_{1/2}$ is illustrated in the inset, which 
plots the time-dependence of $d$, the depth of the 
soliton normalized to its initial value. Dashed lines show the point at which
$d=0.5$ for the first time, while the dotted lines indicate the points at 
which $d=0.45$
and $d=0.55$. These enable us to add error bars to our half-life estimates,
in a manner which could account for experimental uncertainties in the contrast
of the absorption images. Note that these error bars may be asymmetrical due
to the non-monotonic change of the soliton depth arising from soliton-sound 
interactions.

The main graph in Fig.\ \ref{fig:halflife-ins} plots the temperature 
dependence of the half-lives for different initial soliton speeds. These
reveal that the damping time decreases both with increasing temperature, 
as expected, and increasing initial speed. The 
half-life is additionally found to decrease when collisions are taken into 
account, consistent with the energy decay observed in Fig.\ 
\ref{fig:sol_energy}. The latter change is dramatic for low 
temperatures and small initial velocities, although the effect of collisions 
becomes less important at higher temperatures. In particular, at high $T$ and 
large initial $v$, the half-life is 
less than the period of the soliton oscillation (marked with a dashed line) 
even when collisions are absent, indicating that their inclusion
will do little to further enhance the decay. Note that the 
large soliton damping observed here at relatively
low temperatures contrasts with the case of collective excitations,
where finite $T$ effects only become important at higher 
temperatures, $T>0.5\, T_c$ \cite{jackson02b}.

This strong damping effect is present even though we have so far considered
a rather idealized scenario with a large trap aspect ratio. In 
experiments the damping may be even more pronounced. To illustrate 
this, we explicitly consider the parameters of the experiment of Burger 
{\it et al.\ }\cite{burger99}:
$N=1.5 \times 10^5$ $^{87}{\rm Rb}$ atoms, $\omega_z=2\pi\times 14\, {\rm Hz}$,
$\omega_{\perp}=2\pi\times 425\, {\rm Hz}$, and $T \simeq 0.5 \, T_c$
\cite{muryshev02}.

Experimentally, a ``phase-imprinting'' method was used to generate solitons
\cite{burger99}, where light was shone on one half of the condensate to create 
a phase imbalance, $\Delta \phi$. To simulate this, we multiply our 
condensate wavefunction at $t=0$ by 
$e^{i\phi(z)}$, where $\phi(z)=0$ for $z<-\ell_e/2$, and $\phi(z)=\Delta \phi$
for $z>\ell_e/2$. Experimentally, due to diffraction of the light, there is a 
smooth transition of the phase between the two halves, which takes place
over a length of $\ell_e$. We represent this transition with a linear ramp,
$\phi=(z/\ell_e+1/2)\Delta \phi$, for 
$-\ell_e/2<z<\ell_e/2$. 

A further consideration when
simulating the experiment is that the resulting solitons were too small to 
detect {\it in situ}, so the condensate was released from the trap and 
allowed to expand before imaging. We simulate this procedure by allowing the 
system to evolve in the trap for a variable time after phase imprinting, and
subsequently setting the trap potential $V=0$. The column density of 
the condensate is calculated 
$t=4\, {\rm ms}$ later, which simulates a typical absorption image 
obtained experimentally.

Fig.\ \ref{fig:expcol} presents our computed expansion images for 
$\ell_e = 1\, \mu{\rm m}$ and $\Delta \phi = \pi$, for the cases of 
$T=0$ (top) and $T \simeq 0.5\, T_c$ (bottom). The corresponding experimental
images for the first two columns (i.e.\ for times $t \leq 12\, {\rm ms}$) 
were presented in Ref.\ \cite{burger99}, where it was argued that the 
soliton had damped sufficiently to prevent it from being observed at 
subsequent times. On this timescale, our simulated expansion images at finite 
$T$ reveal that the soliton decays to approximately half of its original depth,
in agreement with experiment. Moreover, in its 
subsequent evolution, the soliton spends a large amount of time near
the edge of the trap where it is heavily damped. In fact, our finite 
temperature simulations reveal that the soliton never
actually re-emerges from the edge of the trap, in stark 
contrast to the corresponding prediction of the GP equations (top row of 
Fig.\ \ref{fig:expcol}). Given the limited experimental visibility at the 
edge of the expanded profiles, our findings are thus in good
agreement with the experiment of Ref.\ \cite{burger99}, as well as with the 
estimates of Ref.\ \cite{muryshev02}.

To conclude, we have studied the dynamics of dark solitons in elongated
Bose Einstein condensates using finite temperature simulations.
Unlike the undamped soliton oscillations predicted by the Gross-Pitaevskii 
equation, dissipation arises at finite temperatures, leading to an increase 
in the oscillation amplitude and a decrease in both the soliton depth and 
energy. This dissipation 
increases with temperature and initial velocity, and is in general sensitive
to the inclusion of binary collisions between the atoms. Nevertheless, 
our work indicates that the predicted dark soliton oscillations should
be observable
in realistic elongated geometries, provided that the temperature is a small
fraction of the critical temperature. Direct comparison of our simulations
with the experiment of Burger {\it et al.\ }\cite{burger99} shows very
good agreement, demonstrating that thermal dissipation fully accounts for the
absence of soliton oscillations in this experiment.

This research was supported by EPSRC.

\end{document}